# Fast generation of quantum dynamics data using a GPU implementation of the time-dependent Schrodinger equation

Rei Nagaya[1,2], Haruki Omatsu[3,2], Daniel M. Packwood[2*]

1. Graduate School of Informatics, Kyoto University, 606-8501, Japan
2. Institute for Integrated Cell-Material Sciences (iCeMS), Kyoto University, 606-8501, Japan
3. Graduate School of Engineering, Kyoto University, 615-8510, Japan

* Corresponding author. dpackwood@icems.kyoto-u.ac.jp

**Abstract**

Efficient methods for generating samples of wave packet trajectories are needed to build machine learning models for quantum dynamics. However, simulating such data by direct integration of the time-dependent Schrodinger equation can be demanding, especially when multiple spatial dimensions and realistic potentials are involved. In this paper, we present a graphics processor unit (GPU) implementation of the finite-difference time-domain (FDTD) method for simulating the time-dependent Schrodinger equation. The performance of our implementation is characterized in detail by simulating electron diffraction from realistic material surfaces. On our hardware, our GPU implementation achieves a roughly 350 times performance increase compared to a serial CPU implementation. The suitability of our implementation for generating samples of quantum dynamics data is also demonstrated by performing electron diffraction simulations from multiple configurations of an organic thin film. By studying how the structure of the data converges with sample sizes, we acquire insights into the sample sizes required for machine learning purposes.

**1. Introduction**

New directions in machine learning research emerge from breakthroughs in our ability to acquire data. A prominent example is the area of materials informatics, which emerged from high-throughput implementations of density functional theory (DFT) [1]. Applications of machine learning in other areas of physics, such as microscope image recognition and automated materials fabrication, can also be traced to advancements in data collection in those areas [2 - 4]. Recently, an interesting new direction for machine learning research has been receiving attention: the use of neural networks and Gaussian processes to solve differential equations [5 - 8]. While this area has focused on simple cases so far, its potential for impact is large considering the heavy computational requirements for simulating realistic physical systems by direct integration. In this work, training data must be first obtained by solving the target differential equation at a sample of points in its domain, a hugely demanding task for cases involving multiple spatial dimensions, time, and realistic potentials. In order to support the development of this area, efficient new ways to simulate partial differential equations are needed.

The time-dependent Schrodinger equation (TDSE) is an equation of motion for a quantum

particle. In essence, it describes the propagation of a wave packet in a medium. The TDSE can be written as

$$\frac{\partial \psi(\mathbf{r},t)}{\partial t} = -\frac{i}{\hbar} H \psi(\mathbf{r},t) \quad (1)$$

where $\mathbf{r}$ is the position vector, $t$ is time, $\psi(\mathbf{r}, t)$ is the wave packet, $i$ is the imaginary unit, $\hbar$ is the reduced Planck constant, and $H$ is the Hamiltonian operator, which contains kinetic and potential energy terms. The TDSE has several important applications in materials science. Simulations of the TDSE and related equations can facilitate the interpretation of experimental data obtained by electron beams or other quantum probes of material structure [9 - 12]. The diffusion Monte Carlo method, which computes the ground-state properties of time-independent systems, is based on the TDSE [13, 14]. The time-dependent Kohn-Sham equation, which is closely related to the TDSE, is fundamental to time-dependent DFT, an important technique for simulating the excited state properties of materials [15, 16]. However, the TDSE is not easy to simulate. On the one hand, it has a complex coefficient which necessitates the use of special numerical schemes which couple the real and imaginary parts. On the other hand, for realistic materials these simulations involve three spatial dimensions plus time, as well as potentials which are highly singular in the neighborhood of nuclei and other points. This factor, as well as the unavoidable self-interference effects arising from the simulation boundaries, requires the use of fine meshes and small time steps, usually set on the basis of tedious trial-and-error. The ability to solve the TDSE on the basis of neural networks or Gaussian processes would dramatically facilitate our ability to apply the TDSE in materials science and other areas. However, in order to generate sample data for training such models, efficient and accurate methods of simulating the TDSE are required.

Efficient simulations of partial differential equations require a combination of fast numerical methods and clever hardware implementations. For the case of the TDSE, several numerical methods are widely used. These include explicit schemes involving iterative applications of the time-evolution operator (e.g., [17 - 19]), the finite-difference time-domain (FDTD) method (e.g., [20 - 26]), and others (e.g., [27]). However, while numerical methods for the TDSE have undergone considerable development, less effort has been made to implement these methods in parallel environments or on special hardware such as graphics processing units (GPUs). A GPU implementation of the FDTD method for the TDSE using the CUDA (Computed Unified Device Architecture) package was presented in reference [28]. This implementation achieved speed-ups in the order of 100 times compared to a CPU implementation for the simulations involving simple systems such as hydrogen atoms and harmonic oscillators. A GPU implementation of a generalized FDTD method for non-linear TDSEs was reported in reference [29], which was also validated for relatively simple 1D and 2D systems. Given the immense development and investment that GPU technology is currently receiving as part of the machine learning boom [30], as well as the pressing need to quickly simulate quantum dynamics for materials science applications, further GPU implementations of the TDSE are desirable.

In this paper, we report a new GPU implementation of the FDTD method for simulating the TDSE. We develop and benchmark our method for a realistic and important scenario in materials science: low-energy electron diffraction from a material surface. In experimental materials science, electron diffraction is used to determine the atomic structure of a material surface [31]. Interpretation of experimental diffraction patterns usually requires comparison with electron diffraction simulations [32], making it essential that the TDSE can be simulated accurately and quickly for many different candidate surface configurations. In a previous study, we applied a CPU-implementation of the FDTD method to simulate electron diffraction from a copper surface and the surface of an organic thin film [33]. The GPU implementation in the current paper achieves a 350 times speed-up over the (serial) CPU implementation when benchmarked for copper surface case. To demonstrate the potential for our GPU implementation to generate large samples of quantum dynamics (wave packet) data in a high-throughput manner, we apply it to nearly 900 candidate structures for the organic thin film described above. Furthermore, we draw insights into the sample size requirements for machine learning purposes by exploring how the global and local structure of the sample converges as a function of sample size. This work therefore opens the way towards dataset generation for training neural network or Gaussian process regression models to solve the TDSE and related equations in physics.

This paper is organized as follows. Section 2 presents the FDTD method and our GPU implementation. Section 3 presents benchmarks of our implementation for realistic simulations of electron diffraction from materials surfaces, and shows how it can be used to generate samples of quantum dynamics data. Discussion and conclusions are left to section 4.

## 2. Methods

*2.1. Finite-difference time-domain (FDTD) simulation for electron diffraction*

Low-energy electron diffraction (LEED) is an experimental technique for characterizing the atomic structures of crystalline surfaces. During a LEED experiment, a beam of electrons with kinetic energy in the range of 10 ~ 300 eV is directed towards the surface of a crystalline material. At these low energies, the electrons do not penetrate deeply into the material, and their interactions are mainly limited to the atoms near the material's surface. A small fraction of the electrons undergo diffraction from these atoms, and the resulting diffraction pattern is observed by a detector positioned behind the electron beam source.

In our work, we consider a variant of a LEED experiment in which the electron beam consists of a single short pulse. The pulse is simulated by integrating the time-dependent Schrodinger equation (TDSE) for the electron pulse. Namely,

$$i\hbar \frac{\partial \psi(\mathbf{r},t:\mathbf{q})}{\partial t} = -\frac{\hbar^2}{2m}\nabla^2 \psi(\mathbf{r},t:\mathbf{q}) + V(\mathbf{r}:\mathbf{q})\psi(\mathbf{r},t:\mathbf{q}), \qquad (2)$$

where $\mathbf{r}$ is position, *t* is time, $\mathbf{q}$ denotes a vector of parameters describing the potential

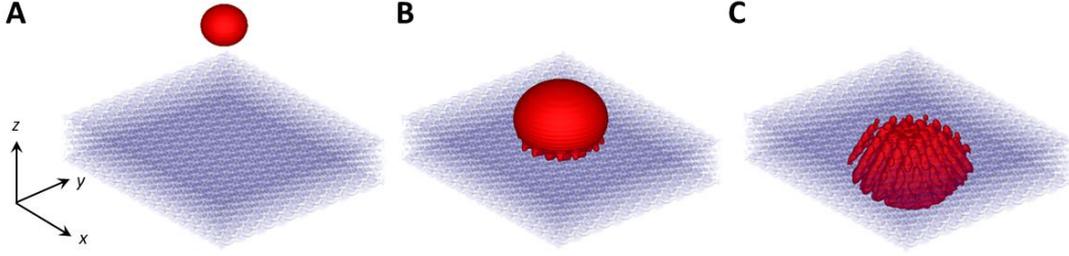

**Figure 1.** Snapshots of a simulation of an electron wave packet propagating towards a copper surface at time (A) 0.20 fs, (B) 0.63 fs, and (C) 0.92 fs. Red shows the wave packet square modulus (95 % electron density isosurface). Blue shows the electrostatic potential of the copper surface (potential within the 95 % percentile). See section 2.1 for simulation details.

(positions and types of the surface atoms, etc), $\psi(\mathbf{r}, t : \mathbf{q})$ is the wave packet describing the electron pulse, $\nabla^2$ is the Laplacian operator, $V(\mathbf{r} : \mathbf{q})$ is an electrostatic potential arising from the surface atoms, $i = \sqrt{-1}$, $\hbar$ is the reduced Plank constant, and $m$ is the electron mass. In the following, we drop explicit reference to $\mathbf{q}$ in our notation unless it is necessary. We simulate equation (2) for $\mathbf{r}$ contained in rectangular simulation cell centered at the origin with periodic boundary conditions. Representative snapshots of such a simulation are shown in Figure 1. The electrostatic potential is calculated prior to the simulation using density functional theory (DFT) as described later, with the crystal surface aligned with the $xy$ plane and the bulk of the crystal terminated after a few atom layers. For the initial condition a Gaussian wave packet is used:

$$\psi(\mathbf{r}, t=0) = \frac{1}{\sqrt{2\pi\sigma^2}} \exp\left(-\frac{|\mathbf{r}-\mathbf{r}_0|^2}{2\sigma^2}\right) \exp\left(-\frac{2i\pi(r_z - r_z^0)}{\lambda}\right), \tag{3}$$

where $\mathbf{r}_0$, $\sigma$, and $\lambda$ indicate the center, spread, and wavelength of the wave packet, respectively. $r_z$ and $r_z^0$ indicate the $z$-components of $\mathbf{r}$ and $\mathbf{r}_0$, respectively. The wavelength is given by

$$\lambda = \sqrt{2\hbar^2\pi^2/mE}, \tag{4}$$

where $E$ is the kinetic energy of the wave packet, which is fixed at the beginning of the simulation. According to equation (3), the wave packet propagates along the $z$ axis in the negative direction towards the surface.

Equation (2) is solved using the finite-difference time-domain (FDTD) technique [21]. In this technique, the wave packet is partitioned into a real component and an imaginary component:

$$\psi(\mathbf{r},t) = \psi_R(\mathbf{r},t) + i\psi_I(\mathbf{r},t). \tag{5}$$

Using the standard rectangular mesh with grid spacings $\Delta x$, $\Delta y$, and $\Delta z$, and also discretizing time with a time step $\Delta t$, these components are then expressed as

$$\psi_R(\mathbf{r},t) = \psi_R^n(i,j,k), \tag{6}$$

and

$$\psi_I(\mathbf{r},t) = \psi_I^n(i,j,k), \tag{7}$$

respectively, where the index $n$ refers to the time point corresponding to $t$ and the indices $(i, j, k)$ refer to the grid location corresponding to point $\mathbf{r}$ (such that $\mathbf{r} = (i\Delta x, j\Delta y, k\Delta z)$ and $i, j, k \geq 0$). The FDTD scheme is then given by:

$$\begin{aligned}
\psi_R^{n+1}(i,j,k) &= \psi_R^n(i,j,k) \\
&\quad - c_x \left( \psi_I^{n+1/2}(i+1,j,k) - 2\psi_I^{n+1/2}(i,j,k) + \psi_I^{n+1/2}(i-1,j,k) \right) \\
&\quad - c_y \left( \psi_I^{n+1/2}(i,j+1,k) - 2\psi_I^{n+1/2}(i,j,k) + \psi_I^{n+1/2}(i,j-1,k) \right) \\
&\quad - c_z \left( \psi_I^{n+1/2}(i,j,k+1) - 2\psi_I^{n+1/2}(i,j,k) + \psi_I^{n+1/2}(i,j,k-1) \right) \\
&\quad + c_v V(i,j,k) \psi_I^{n+1/2}(i,j,k),
\end{aligned} \tag{8}$$

and

$$\begin{aligned}
\psi_I^{n+1/2}(i,j,k) &= \psi_I^{n-1/2}(i,j,k) \\
&\quad + c_x \left( \psi_R^n(i+1,j,k) - 2\psi_R^n(i,j,k) + \psi_R^n(i-1,j,k) \right) \\
&\quad + c_y \left( \psi_R^n(i,j+1,k) - 2\psi_R^n(i,j,k) + \psi_R^n(i,j-1,k) \right) \\
&\quad + c_z \left( \psi_R^n(i,j,k+1) - 2\psi_R^n(i,j,k) + \psi_R^n(i,j,k-1) \right) \\
&\quad - c_v V(i,j,k) \psi_R^n(i,j,k),
\end{aligned} \tag{9}$$

where $c_x = \hbar\Delta t/(2m\Delta x^2)$, $c_y = \hbar\Delta t/(2m\Delta y^2)$, $c_z = \hbar\Delta t/(2m\Delta z^2)$, and $c_v = \Delta t/\hbar$. According to the above, $\psi_R$ is computed at integer-valued time points, and $\psi_I$ is computed at half-integer valued time points. The square amplitude (electron density) at timestep $n$ is then approximated by $|\psi^n(i,j,k)|^2 = \psi_R^n(i,j,k)^2 + \psi_I^{n-1/2}(i,j,k)^2$.

A significant difficulty when simulating wave packet dynamics is the presence of unphysical self-interaction effects. These arise when the wave packet undergoes interference with its image due to periodic boundary conditions, or when the wave packet reflects from the simulation boundaries when reflective boundaries are used. In order to suppress such self-interaction effects, a complex absorbing potential (CAP) was used. To implement the CAP, the following potential

$$U(i,j,k) = \begin{cases} (-ie_c\alpha/\Delta z)(r_{cut} - k\Delta z)^2 & k\Delta z < r_{cut} \\ 0 & \text{otherwise} \end{cases} \tag{10}$$

was added to the electrostatic potential $V(i, j, k)$ at each mesh point $(i, j, k)$, where $\alpha$ is a constant and $e_c$ is the electron charge [34, 35]. According to equation (10), the CAP is active in the regions where $z < r_{cut}$, i.e., at the bottom of the simulation domain.

Figure 1 shows snapshots of the electron pulse proceeding towards a copper(111) (Cu(111)) surface, as simulated according to the FDTD method above using a spatial domain of 40.9 x 40.9 x 70 Å and a spatial grid size of $N_x$ x $N_y$ x $N_z$ = 240 x 240 x 420. The time step was set following the recommendation in [31]:

$$\Delta t = \frac{m}{5\hbar} \min(\Delta x, \Delta y, \Delta z)^2. \tag{11}$$

This works out to be $\Delta t$ = 0.38 as for the settings described above. The surface was modelled as a four-layer slab of copper atoms duplicated by 15 x 15 unit cells in the $x$ and $y$ direction. The electrostatic potential was calculated from DFT. Detailed parameter settings are given in section 2.3.

*2.2. GPU implementation of the FDTD method*

The computational demands of the FDTD scheme will clearly increase as the number of grid points and time steps increases. In order to accelerate the FDTD scheme, we implement it within the highly parallel, multicore architecture of a graphics processing unit (GPU). A GPU consists of multiple cores arranged into several streaming multiprocessors (SM). Each SM contains 32 cores, as well as a shared memory unit for the cores, and other units for loading and storing data, scheduling, and for executing transcendental functions.

Our implementation utilizes NVIDIA's CUDA (Compute Unified Device Architecture) library for the C++ language. A program which uses CUDA should follow a hierarchical structure (Figure 2A). At the highest level of this hierarchy is the so-called *grid*. The grid in turn contains several *blocks*. Each block in turn contains multiple *threads*, which contain individual instructions from the code. In terms of hardware, the grid, blocks, and threads roughly correspond to the GPU itself, the SMs, and the individual cores. A *warp* is a group of threads which are executed on a single GPU core. At runtime, the threads are bundled into warps, which are then assigned to cores within each SM. Usually, 32 threads are assigned to a single warp. The execution of the program is controlled by a so-called kernel function, which is called from CPU code and executes the program by assigning different parts of it to each thread.

Within CUDA, each block is identified by two indices ($b_X$, $b_Y$). The threads within a block are identified by three indices ($t_X$, $t_Y$, $t_Z$). Thus, to implement the FDTD scheme within CUDA, a mapping between the grid indices ($i, j, k$) of the spatial domain and the block and thread indices is required. We implement this mapping in several steps (Figure 2B). In the first step, the spatial domain is partitioned into multiple cuboidal regions, each containing $B_x \times B_y \times B_z$ grid points. Each plane of cuboids parallel to $xz$ are then laid side-by-side in a two-dimensional array, beginning with the plane touching $y$ = 0 and proceeding with subsequent planes in order of increasing *y* coordinate. This new two-dimensional array defines the CUDA grid, and each cuboid defines a CUDA block. CUDA threads correspond to the spatial domain grid points contained within each cuboid. For the new two-dimensional array, the long axis (the direction along which the planes of cuboids were laid) is denoted *Y* and the short axis is denoted *X*. In our scheme, the mapping between the spatial grid indices and the CUDA block and thread indices is therefore

$$i = \left(b_Y B_y + t_Y\right) \bmod N_x$$
$$j = \left\lfloor \left(b_Y B_y + t_Y\right)/N_x \right\rfloor \times B_z + t_Z \qquad (12)$$
$$k = b_X B_x + t_X.$$

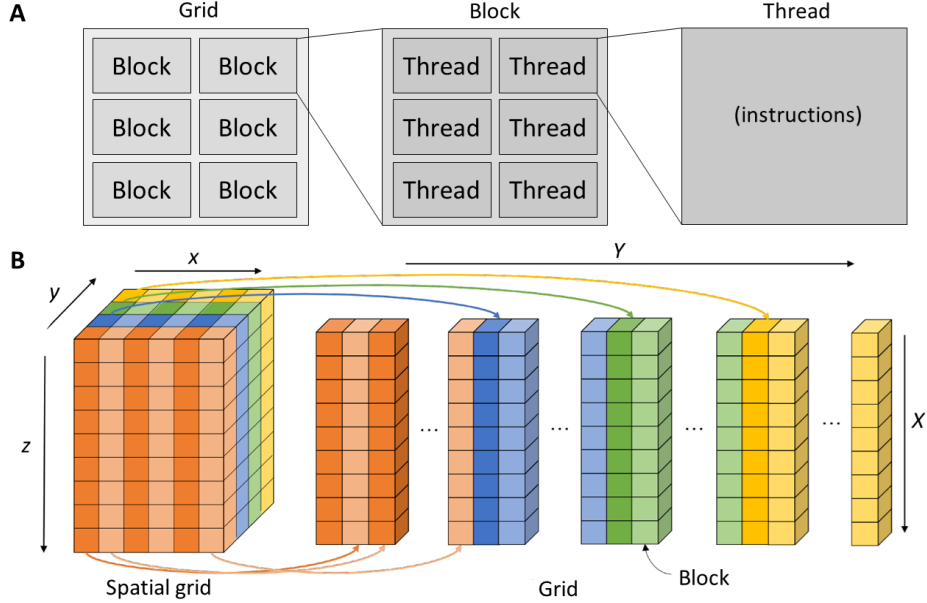

**Figure 2.** (A) Sketch of the hierarchical structure of a CUDA program. (B) Correspondence between the spatial grid points and CUDA elements in our implementation. See text for details.

In addition to grid point mappings it is also important to consider memory management in order to achieve an efficient GPU implementation. In our implementation, we utilize the *coreless access* feature included in NVIDIA GPUs. Coreless access allows for each wrap to read and write a unique continuous region of the GPU's global memory. In the FDTD method, two objects need to be stored in the global memory in order to compute the wave packet at time step $n+1$, namely $\{\psi_R^n(i,j,k)\}$ and $\{\psi_I^{n-1/2}(i,j,k)\}$, the set of all wave function values at all spatial grid points. In our implementation, the spatial grid is converted into a one-dimension vector for storage in the GPU's global memory. This conversion is achieved by looping over the *x*, *y*, and *z* indices of the spatial grid points, in such a way that *z* is the inner-most loop (fastest index) and *x* the outer-most loop (slowest index). This is convenient from a programming perspective, as the correspondence between the *z* axis of the spatial domain and the *X* axis of the CUDA grid can be used to establish a mapping between memory locations and spatial points.

As mentioned above, all threads within a warp are executed simultaneously on the same core at runtime. However, potential performance issues arise from the way in which this simultaneous execution is performed within CUDA. Specifically, for each thread, each line of code is read and executed sequentially. The program moves to the next line of code only once the current line of code has been successfully executed for all threads. This is a problem at 'branching points' (such as at 'if' conditions, at which the thread must satisfy some condition before the program can proceed), as it means that the program must wait until all threads satisfy the conditions of the branch. In the FDTD method, branching points can potentially arise when boundary conditions are implemented. For example, in a typical (CPU) code, periodic boundary conditions are typically implemented as

$$\psi_R^{n+1}(i+1,j,k) = \begin{cases} \psi_R^{n+1}(0,j,k) & i = N_x - 1 \\ \psi_R^{n+1}(i+1,j,k) & \text{otherwise} \end{cases} \tag{13}$$

and

$$\psi_R^{n+1}(i-1,j,k) = \begin{cases} \psi_R^{n+1}(N_x-1,j,k) & i = 0 \\ \psi_R^{n+1}(i-1,j,k) & \text{otherwise} \end{cases}, \tag{14}$$

and similarly for the indices $j$ and $k$, as well as the imaginary component $\psi_I^{n+1/2}$. In our implementation, we avoid branching points by simply replacing $i + 1$ and $i$ -1 in the above expressions with $(i + 1)$ mod $N_x$ and $(i – 1 + N_x)$ mod $N_x$ respectively, and similarly for the other dimensions.

As can be seen from equations (8) and (9), calculation of the wave packet at point $(i, j, k)$ requires access to data from the surrounding points $(i \pm 1, j \pm 1, k \pm 1)$. This means that within each block, data from spatial grid points adjacent to the original cuboidal region will be required. For this reason, we make use of the shared memory of the SMs. Specifically, for each block, we include the data for these adjacent points in the shared memory of the corresponding SM. It should be noted that this treatment introduces branching points in the code (in order to identify when data from these adjacent points should be accessed), introducing a computational time trade-off associated with block numbers and sizes.

*2.3. Simulation details and parameter settings*

In this work, two sets of FDTD simulations were performed. The first set of simulations were performed to evaluate the performance of the GPU implementation. These simulations considered a copper(111) (Cu(111)) surface, which was modelled as a four-layer slab of copper atoms (Figure 3A). The slab was aligned in the *xy* plane and positioned such that the *z* coordinate of the top layer was 28.35 Å. Above the top layer a 41.65 Å vacuum space was added. To create the spatial domain for these FDTD simulations, we first calculated the electrostatic potential *V* from DFT and then duplicated it by 15 x 15 unit cells in the *x* and *y* direction. The resulting spatial domain had dimensions of 35.4 x 40.9 x 70.0 Å. The spatial grid points were then generated by specifying the desired number of divisions along each axis ($N_x$, $N_y$, and $N_z$), and the values of the electrostatic potential at each grid point $V(i, j, k)$ were obtained by interpolating the calculated potential. These simulations considered various spatial domain sizes between $N_x$ x $N_y$ x $N_z$ = 140 x 140 x 420 through to 240 x 240 x 420. For each simulation, a CAP with $\alpha = 10^8$ eV and $r_{cut} = 0$ was incorporated by appending a region with thickness of 100 grid points in *z* direction to the bottom of the spatial domain (see equation (10)). In each simulation, the initial wave packet in equations (3) and (5) was positioned in the center of the *xy* plane with a vertical position of $r^0_z = 42.0$ Å, and with spread parameter $\sigma = 1.8$ Å and kinetic energy parameter $E = 60$ eV. All simulations were performed for 8000 time steps. To test the effect of block sizes in our CUDA implementation, we examined the cases of $B_x$ x $B_y$ x $B_z$ of 16 x 4 x 2, 16 x 4 x 4, 16 x 8 x 4, and 16 x 8 x 8.

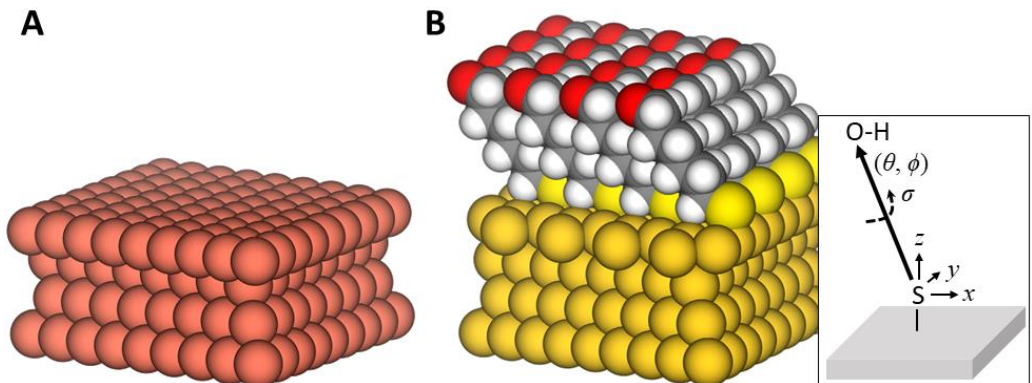

**Figure 3**. Atomistic slab models of the surfaces used in our simulations. (A) Cu(111) slab. (B) Au(111)-C$_6$SAM slab. Slabs have been expanded into 4 x 4 supercells for clarity. Coral-colored spheres = Cu atoms, dark yellow spheres = Au atoms, bright yellow spheres = S atoms, grey spheres = C atoms, white spheres = H atoms, red spheres = O atoms. The insert in (B) shows the variables used to describe the orientation of the C$_6$SAM molecule. $\theta$ and $\phi$ are respectively the altitude and azimuthal angles of the vector connecting the S and O atoms. The internal orientation $\sigma$ is the angle made between the projection of the O-H bond on the *xy* plane and the *x* axis.

The second set of simulations was used to illustrate how our GPU implementation can be used for dataset generation. These simulations considered electron scattering from a monolayer of the molecule HO(CH$_2$)$_6$S (abbreviated C$_6$-SAM) covalently bonded to a gold(111) (Au(111)) surface through the S atom (Figure 3B). A total of 879 simulations were performed, each with a different configuration for the C$_6$-SAM molecule. To generate each configuration, a four-layer gold slab was built with only a single Au(111) unit cell in the *xy* plane. A single C$_6$-SAM molecule was then attached to the Au atom of the top-most layer, and its orientation was adjusted by setting the following three variables (Figure 3B insert): the azimuthal orientation of the S-O atom axis $\phi$, the elevation of the S-O axis $\theta$, and the rotation angle of the molecule about the S-O axis $\sigma$. Each configuration was generated by sampling $\phi$, $\theta$, and $\sigma$ at random, in such a way that the van der Waals exclusion radii of each atom was not violated. The resulting structure (Au slab and C$_6$-SAM molecule) was then positioned so that the Au atom planes were parallel to *xy* and the *z*-coordinate of the upper-most C$_6$SAM atom was 28.35 Å. Above this atom, a 41.65 Å vacuum space was added. The spatial domains for the simulations were then created by calculating an electrostatic potential using DFT and duplicating it as described above. For each simulation, the spatial domains had dimensions 79.9 x 92.3 x 70.0 Å and

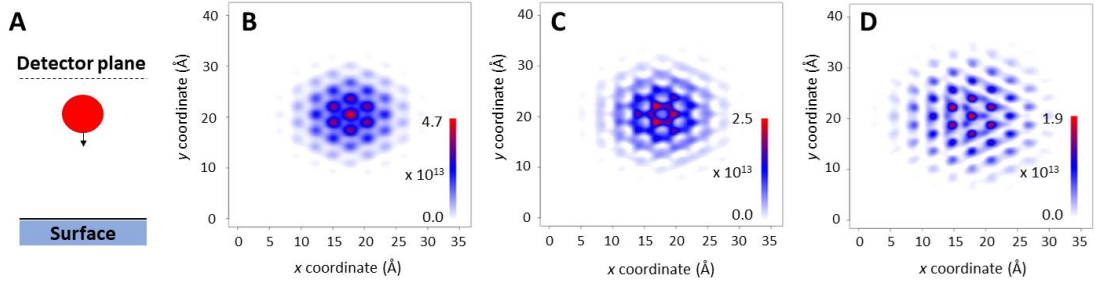

**Figure 4.** (A) Definition of the detector plane. The image is viewed within the *xz* plane. The red ball denotes the wave packet. (B) Snapshots of detector-plane electron density after 0.83 fs for the simulation shown in Figure 1. (C) Detector-plane electron density after 0.87 fs. (D) Detector-plane electron density after 0.92 fs.

240 x 240 x 420 grid points. A CAP region was added to the bottom of each spatial domain using the same settings as above. In each simulation, the initial wave packet was set as described above, but with a spread parameter of $\sigma = 8.0$ Å and kinetic energy $E = 15$ eV. Each simulation was performed for 5000 time steps with the time step set as in equation (11) above. Block sizes for our CUDA implementation were fixed to 16 x 4 x 4 for all simulations.

For both simulations, the electron density arriving at the 'detector' plane, defined as

$$\rho(t) = \{\psi_R^2(\mathbf{r},t) + \psi_I^2(\mathbf{r},t) : \mathbf{r} \in D\}, \tag{15}$$

where $D = \{(r_x, r_y, r_z): r_z = z_{det}\}$ and $z_{det}$ is the detector plane position, was retained for every $h$th time step (Figure 4A). $h$ was set to 20 for all simulations reported here. $z_{det}$ was set to 37.6 Å for the first set of simulations (Cu(111) case) and 45.6 Å for the second set ($C_6$-SAM case). The set $\rho = \{\rho(0), \rho(h\Delta t), \rho(2h\Delta t), \ldots\}$ represents a space-time trajectory for the electron density arising from a fixed configuration of surface atoms (Figure 4B - D). The detector-plane electron density in (15), rather than the entire electron density, was dumped to reduce data storage requirements. For the simulations reported in section 4, a dump of detector-plane electron density for a single time step required only around 670 KB of space, compared to around 300 MB for a dump of the entire electron density.

All DFT calculations for the electrostatic potentials were performed using the VASP code (version 5.4.4 [36]) using PAW-PBE pseudopotentials [37], and 4 x 4 x 1 k-points grids centered at the gamma point. For the first set of simulations, the electrostatic potentials were calculated using the LDA exchange-correlation functional [38] and a 550 eV basis set cut-off. For the second set of simulations, the rev-vdW-DF2 exchange-correlation functional [39] and a 650 eV basis-set cut-off was used.

*2.4. Hardware details*

Simulations were performed on a workstation equipped with an Intel Xeon E5-2603 v3 1.6 GHz core and a NVIDIA Quadro K620 GPU, and running an Ubuntu 18.04.06 LTS operating system. All codes were written in C++ and compiled using g++ version 7.5.0 and CUDA version nvcc 11.3. We also compare our GPU implementation to an OpenMP parallelized CPU implementation, which was ran on multicore server equipped with 16 double-threaded Intel Xeon Gold 6242 2.8 GHz cores and running CentOS Linux 7. This multicore server was also used for all DFT calculations.

## 3. Results and discussion

### 3.1. Performance of GPU implementation

The first set of simulations described above were used to evaluate the performance of our GPU implementation of the FDTD method. Figure 5 compares the timestep-averaged execution time of our GPU implementation (data marked 'GPU') to a serial code running on a CPU ('CPU') and a parallelized core running on a multicore server ('cluster'). The data marked 'transfer' indicates the time required to transfer data (potential energy data, grid spacings, and other parameters) to the GPU, which occurs before the start of the FDTD iterations. In all of these simulations, the block size $B_x$ x $B_y$ x $B_z$ was fixed at 16 x 4 x 4. Mesh sizes $N_x$ x $N_y$ x $N_z$ were varied between 140 x 140 x 420 and 240 x 240 x 420, which correspond to a total of 10192000 and 2995200 mesh points, respectively (including the additional mesh points in the CAP region).

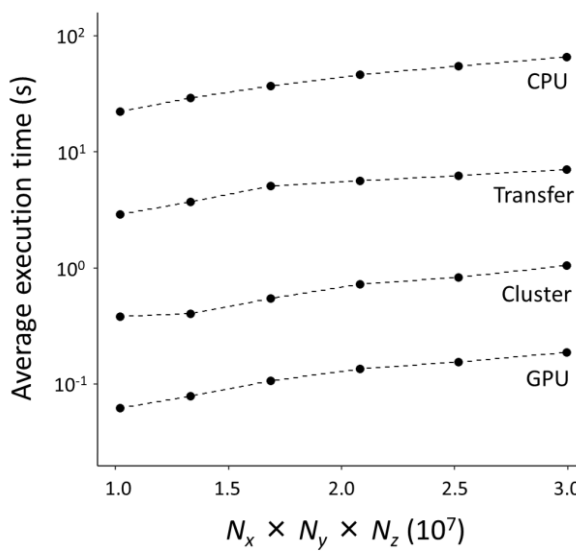

According to Figure 5, the GPU implementation is by far the most efficient implementation of the FDTD method. For all mesh sizes tested, the GPU implementation is roughly 350 times faster than the serial CPU implementation and 5.5 times faster than the multicore CPU implementation. The latter result is particularly significant considering the overall hardware superiority of our multicore server compared to the old workstation on which the GPU calculations were performed. Calculation times increase linearly with grid sizes. A linear regression analysis of the execution time shows that the GPU calculation time increases by around 6 ns per mesh point, compared to 2000 ns for the serial CPU implementation and 30 ns for the cluster implementation.

**Figure 5**. Timestep-averaged execution times for our GPU implementation (GPU) compared to serial implementation (CPU) and an implementation on high-performance server (Cluster; see text for hardware details). 'Transfer' refers to the CPU-to-GPU data transfer time, which occurs prior to the start of the simulations. The horizontal axis is the total number of points in the spatial grid.

The calculations times for the GPU case discussed above were obtained by averaging over all FDTD iterations. To discuss the contribution to the overall code execution time of the data transfer process (which occurred prior to the FDTD steps), we consider the line denoted 'Transfer' in Figure 5. It is clear that the data transfer process is very slow, requiring around 40 times as much computational time compared to an average FDTD iteration. The large times associated with data transfer represent an overhead to our simulations, but do not reflect on the performance of our FDTD implementation itself. Even if data transfer times are considered, the performance of our GPU implementation remains 50 – 100 times faster than the ordinary serial implementation.

|  | Without coreless access | With coreless access | | |
|---|---|---|---|---|
| **Execution time (s)[a]** | 0.52 | 0.16 | | |
|  | Without shared memory | With shared memory | | |
| **Execution time (s)[b]** | 0.16 | 0.20 | | |
|  | **Block size ($B_x \times B_y \times B_z$)** | | | |
|  | $16 \times 4 \times 2$ | $16 \times 4 \times 4$ | $16 \times 8 \times 4$ | $16 \times 8 \times 8$ |
| **Execution time (s)[c]** | 0.17 | 0.16 | 0.17 | 0.21 |

Table 1. Timestep-averaged execution times for various settings of our GPU implementations. All simulations used mesh sizes of 240 × 240 × 420. (a) Simulations using block sizes of 16 × 4 × 4 without shared memory. (b) Simulations with coreless access and block sizes of 16 × 4 × 4. (c) Simulations using coreless access and without shared memory.

In addition to the comparison above, we also explored the effect of coreless access, GPU shared memory, and block sizes on our GPU implementation. Coreless access refers to the reading and writing of continuous global memory from the GPU. Within the CUDA platform, coreless access can be performed per warp. In the simulations above, the coreless access option was included. To evaluate the effect of coreless access, we performed simulations for 100 timesteps using a mesh size of 240 x 240 x 420 and block sizes of 16 x 4 x 4, and computed the average computational time per time step with and without this option. Without the use of coreless access, the average computation time became 0.52 s (Table 1). This compares to an average time of around 0.16 s when coreless access is included. This significant decrease in computational time suggests that coreless access should be used for efficient simulations with our implementation.

In the simulations described above, the shared memory option was not used. To test its effect on simulation times, we performed another set of simulations for 100 time steps using the same mesh and block sizes described above, and again calculated the average computation time per step. With the use of shared memory, the average computational time rose slightly to 0.20 s, which compares to 0.16 s without the use of shared memory (Table 1). The minor increase in computational time therefore suggests against the use of shared memory in our implementation.

In a similar manner as described above, we also tested the effect of block sizes. With mesh size fixed at 240 x 240 x 420, coreless access included, and no use of shared memory, the time step-averaged computational was calculated for the cases $B_x$ x $B_y$ x $B_z$ = 16 x 4 x 2, 16 x 4 x 4, 16 x 8 x 4, and 16 x 8 x 8. For these cases, we obtained 0.17 s, 0.16 s, 0.17 s, and 0.21 s, respectively (Table 1). Among the cases tested, the block size of 16 x 4 x 4 was found to be most efficient. However, the variation between these computational times is small, suggesting that the sensitivity of our results to block sizes is not large.

*3.2. Trajectory sample generation and sample size dependence*

Having established the superior efficiency of our GPU implementation of the FDTD method, we now illustrate how it can be used to generate samples of quantum dynamics data. We generated 879 configurations (orientations) of the $C_6$-SAM monolayer adsorbed to Au(111) and simulated electron diffraction from each one. The result is a sample of 879 trajectories of the detector-plane electron density, each obtained for a different realization of the parameter vector **q** (for the present case, **q** consists of the orientation angles $\theta$, $\phi$, and $\sigma$ shown in Figure 3B). In a machine learning application, such a sample would serve as training data for fitting a regression model to predict $|\psi(\mathbf{r}, t : \mathbf{q})|^2$ for any space-time coordinate (**r**, *t*) and **q**.

We first evaluate the time required to obtain this sample. For each simulation, three distinct steps are involved: (i) computation of the electrostatic potential for a single unit cell of the Au(111)-$C_6$SAM system from DFT, (ii) expansion and interpolation of the electrostatic potential onto the mesh points of the spatial domain, and (iii) execution of the FDTD simulation on the GPU. These steps are not directly comparable due to the different hardware used for each case (a multicore server in (i), a single core from a workstation in step (ii), and a GPU in step (iii)). Nonetheless, we report average computational times to provide a sense of the effort required to generate this database. Moreover, the hardware that we used are typical of the hardware available to academic research groups. The average computational time required for steps (i) and (iii) for a single $C_6$SAM configuration is around 4 mins and 2.4 mins, respectively, using the hardware described above. These computational times are quite tolerable for the purpose of generating a database of the size used here. However, more problematic is step (ii), which required around 15 mins for our case. This step was performed using a combination of an R script to read and write data as well as a C++ code compiled with the nearest neighbor search library ANN to interpolate the electrostatic potential [40, 41]. The inefficiency of this step is mainly attributed to the use of a high-level programming environment (R) to read and write the large mesh point coordinate data, the use of only one processor, as well as the highly unoptimized R script itself. At the time of the research these codes were not designed for optimal performance, and therefore lower computational times for step (ii) should be possible. Nonetheless, these results show that it is the interpolation step, and not the electrostatic potential calculation nor the GPU-based FDTD simulation, which make the decisive contribution to the computational times required for sample generation.

For machine learning applications, the required sample size should be considered in addition to the sample generation times. In other words, it is important to examine how the sample data – represented as a point cloud in an appropriate space – converges to a limiting distribution as sample size increases. There are two ways in which a point cloud representation of our data could be obtained. The first is by proposing a feature representation (descriptors) for the data, the components of which would serve as spatial coordinates. The second way is propose a measure of distance between pairs of sample points, and then embed the sample into a low-dimensional manifold in such a way that the distance between pairs is preserved. We adopt the second approach here, because it is beyond it is beyond the scope of this work to design descriptors for wave packet trajectory data. Thus, we compute the following dissimilarity metric between every pair of trajectories $k$ and $j$ in our sample:

$$d_{kj} = \frac{1}{t_1 - t_0} \int_{t_0}^{t_1} \left( \int_D \left( |\psi(\mathbf{r},t:\mathbf{q}_k)|^2 - |\psi(\mathbf{r},t:\mathbf{q}_j)|^2 \right)^2 d\mathbf{r} \right)^{1/2} dt, \quad (16)$$

where $\mathbf{q}_k$ represents the parameter vector for trajectory $k$ and $D$ denotes the detector plane. Equation (16) therefore compares trajectories according to the pointwise overlap of their detector-plane electron density, averaged over the time window $t_1 - t_0$. In this work, we set $t_0 = 1.24$ fs and $t_1 = 2.39$ fs. The UMAP technique was then used to obtain a two-dimensional point cloud representation of our sample. In order to highlight the local structure of the point cloud, we employed the Partition Around Medoids (PAM) technique to partition the sample into clusters. The number of clusters to detect was set to 15, justified on the basis of a previous study which identified 12 clusters for the same system using a much smaller dataset [33]. The UMAP technique was performed using the implementing in the R packages umap and Mercator [42, 43].

The resulting point cloud is shown in Figure 6. Kernel density estimation from the R package MASS [44] was applied to help visualize the overall distribution of trajectories in the sample more clearly (black background). It can be seen that the distribution is quite heterogeneous, consisting of a patchy, multi-modal structure with a mixture of high- and low-density regions. The assignment of trajectories into the clusters appears reasonable, with each of the 15 clusters mostly occupying different regions of the distribution. One of the clusters is very large, containing 286 trajectories, whereas the others only contain between 18 and 84 trajectories.

We explore sample size sufficiency by two approaches: through convergence of global data structure and through the convergence of local data structure. In order study global structure convergence, we employ the following statistic. For a fixed sample of trajectories of size $n$, define the *average cluster dispersion* as

$$acd = \frac{1}{m} \sum_{i=1}^{m} \left[ (\overline{x}_i - \overline{x})^2 + (\overline{y}_i - \overline{y})^2 \right]^{1/2}, \quad (17)$$

where $m$ is the number of clusters ($m = 15$), $(\overline{x}, \overline{y})$ is the coordinates of the mean of the entire sample, and $(\overline{x}_i, \overline{y}_i)$ is the coordinates of the mean of the points in cluster $i$. Thus, (17) measures the average distance of the cluster centers from the sample center. In order to study local structure convergence of our data, we define the *average intra-cluster dispersion*:

$$aicd = \frac{1}{n} \sum_{i=1}^{m} \sum_{j=1}^{m_i} \left[ (x_{ji} - \overline{x}_i)^2 + (y_{ji} - \overline{y}_i)^2 \right]^{1/2}, \quad (18)$$

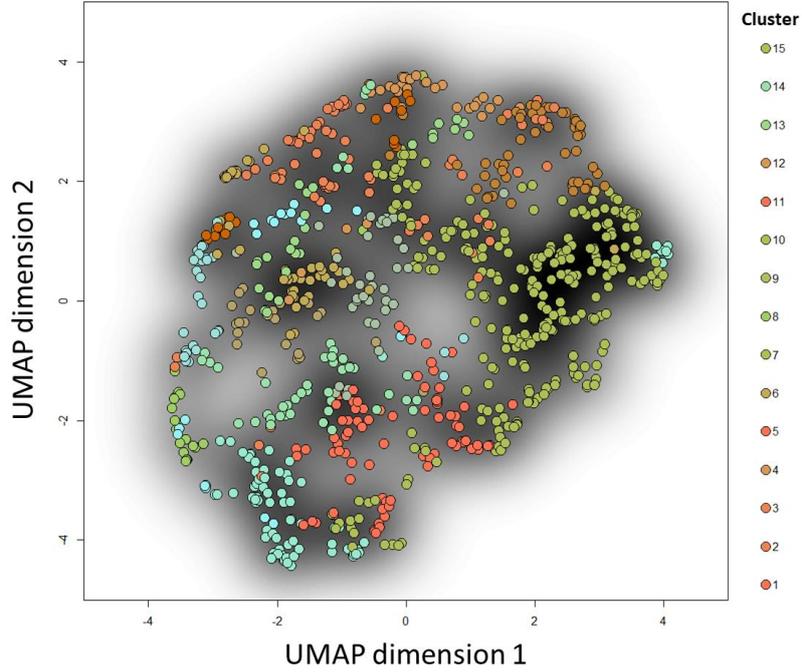

**Figure 6.** Visualization of a database of wave packet space-time trajectories. Each point corresponds to one wave packet trajectory simulated for a fixed parameter vector **q** using our GPU implementation. Visualization used the UMAP method. Points are coloured according to their cluster label. The black background visualizes the overall distribution of points, as estimated with kernel density estimation method with bandwidth = 2.

where $n$ is the size of the sample, $m_i$ the size of cluster $i$, and $(x_{ji}, y_{ji})$ the coordinate of the $j$th point in cluster $i$. Thus, (18) measures the average distance of the points from the centers of their respective clusters. Figures 7 A and B plots the *acd* and *aicd* for various sample sizes. Each point in these plots is an average of 50 independent random subsamples of the original parent sample. The error bars correspond to two times the standard error. Both *acd* and *aicd* increase logarithmically with sample size. The *acd* grows with sample size $n$ through a fast phase up until about $n = 600$, and then through a slower asymptotic phase from about $n = 500$ onwards. Within the range of statistical error, the global structure of the sample appears relatively stable from about $n = 600$ onwards, however there remains a tendency for the *acd* to increase on average. The *aicd* appears to converge more slowly with $n$, suggesting that stable local structures require larger sample sizes.

Averages are not necessarily the most reliable way to monitor sample size convergence, because they are sensitive to outliers. We therefore consider median values as well, which are insensitive to outliers by definition. As an alternative to the *acd*, we therefore consider a *median cluster dispersion* (*mcd*), defined as the median value of terms in the square backets of (17). Likewise, we can define a *median intra-cluster dispersion* (*micd*), defined as the median value of the terms in the square brackets of (18). The *mcd* and *micd* are plotted in Figure 7C and D. Again, the points are an average of median values from 50 independent random subsamples (the error bars correspond to two times $[\pi\sigma^2/(2(n-2))]^{1/2}$, the approximate standard deviation of the sampling distribution of the median [45]). It is clear from Figure 7C that *mcd* converges for about $n = 600$ onwards. This result, taken together with the slower convergence of the *acd*, confirms that the global structure of the sample is stable from about $n = 600$ onwards, save for the few 'outlier' clusters which take a longer time to stabilize as *n* increases. On the other hand, the behavior of *micd* is essentially indistinguishable from the *aicd*, showing that slow local convergence is an intrinsic effect of the data and cannot be attributed to outliers alone. This confirms that larger sample sizes are required to achieve stable local structures.

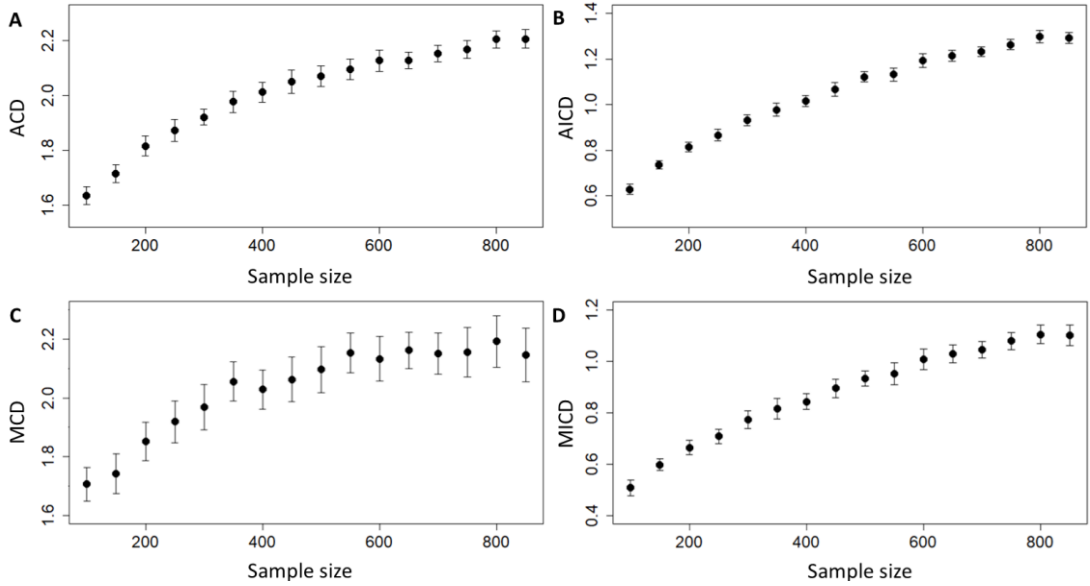

**Figure 7.** Data dispersion statistics and their dependence on sample sizes. ACD = Average cluster dispersion, AICD = average intra-cluster dispersion, MCD = median cluster dispersion, MICD = median intra-cluster dispersion. See text for details.

What do these results say about the prospects of building a kernel ridge regression or neural network model for predicting $|\psi(\mathbf{r}, t : \mathbf{q})|^2$? For the case of such models, the required sample sizes depend upon the choice of feature representation for the wave packet trajectories. However, consider the case where the feature representation is such that the distance metric in (16) is approximately satisfied. Then, these results suggest that relatively small sample sizes (around 600 – 800 wave packet trajectories) would be sufficient for building a model which can reproduce the essential (global) dependence of wave packet trajectories on **q**. However, for very accurate models which can reproduce effects arising from tiny changes in **q**, larger samples in excess of 800 wave packet trajectories may be required.

## 5. Conclusions

Regression models for predicting wave packet evolution require large samples of wave packet trajectory data for training. In this work, we presented a GPU implementation of the finite-difference time-domain method for integrating the time-dependent Schrodinger equation. Our implementation was illustrated by simulating low-energy electron diffraction from copper surfaces and organic thin films. Electron diffraction is important in materials science and also a realistic and challenging target for quantum simulation. However, it is important to emphasize that our GPU implementation is not restricted to this case will apply to any simulation involving a regular 3D grid. For the simulations performed here, our GPU implementation achieved a 350 times speed-up compared to a serial CPU implementation. We demonstrated its use for generating large database of wave packet trajectories for machine learning applications and explored how data structure converges with sample size. Sample sizes in the order of 600 appear sufficient for broadly capturing wave packet dependence on electrostatic potential, however larger sample sizes appear necessary to capture detailed wave packet changes resulting from smaller changes in potential.

Several directions for future research are suggested by this work. The way in which spatial grid points are mapped to threads within the CUDA platform essentially defines the GPU implementation. Further work could therefore be performed to determine the optimal mapping for the case of the time-dependent Schrodinger equation. Overheads arising from data transfer between the CPU and GPU were also found to be significant. Special data transfer strategies should therefore be explored. Finally, for machine learning applications further aspects of sample generation should be considered. Regardless of computational speeds, large samples may be simply impractical, as hundreds of megabytes may be required to store wave packet data arising from a single time step. These space requirements will multiply by the thousands when simulations are performed over long time periods. These space requirements will multiply again when mesh sizes are increased. Efforts to rigorously establish low-dimensional representations of such data (such as detector-plane electron densities, as used in this work), or to develop cluster sampling schemes by exploiting the presence of clustering in the data, might therefore be helpful.

## Acknowledgements

This work was supported by JSPS KAKENHI grant 18K14126 and the Institute for Integrated Cell-Material Sciences.